\documentclass[aps,prx,reprint,superscriptaddress]{revtex4-2}
\usepackage{siunitx}
\usepackage{mhchem}
\usepackage{bm}
\usepackage{graphicx}
\usepackage[colorlinks]{hyperref}
\usepackage{amsmath}

\newcommand{\figurewidthwide}{.80\textwidth}
\newcommand{\figurewidthhalf}{.45\textwidth}

\begin{document}

% Title of paper
\title{Negative electrocaloric effect in nonpolar phases of perovskite over wide range of temperature}

\author{Xingyue Ma}
\affiliation{National Laboratory of Solid State Microstructures and Collaborative Innovation Center of Advanced Microstructures, Nanjing University, Nanjing 210093, China}
\affiliation{Jiangsu Key Laboratory of Artificial Functional Materials, Department of Materials Science and Engineering, Nanjing University, Nanjing 210093, China}

\author{Mingxing Chen}
\affiliation{School of Physics and Electronics, Hunan Normal University, Key Laboratory for Matter Microstructure and Function of Hunan Province, Key Laboratory of Low-Dimensional Quantum Structures and Quantum Control of Ministry of Education, Changsha 410081, People's Republic of China }

\author{Jun-Ming Liu}
\affiliation{National Laboratory of Solid State Microstructures and Collaborative Innovation Center of Advanced Microstructures, Nanjing University, Nanjing 210093, China}

%\author{L. Bellaiche}
%\affiliation{Physics Department and Institute for Nanoscience and Engineering, University of Arkansas, Fayetteville, %Arkansas 72701, USA}

\author{Di Wu}
\email{diwu@nju.edu.cn}
\affiliation{National Laboratory of Solid State Microstructures and Collaborative Innovation Center of Advanced Microstructures, Nanjing University, Nanjing 210093, China}
\affiliation{Jiangsu Key Laboratory of Artificial Functional Materials, Department of Materials Science and Engineering, Nanjing University, Nanjing 210093, China}

\author{Yurong Yang}
\email{yangyr@nju.edu.cn}
\affiliation{National Laboratory of Solid State Microstructures and Collaborative Innovation Center of Advanced Microstructures, Nanjing University, Nanjing 210093, China}
\affiliation{Jiangsu Key Laboratory of Artificial Functional Materials, Department of Materials Science and Engineering, Nanjing University, Nanjing 210093, China}

\date{\today}

\begin{abstract}

The electrocaloric effect (ECE) offers a promising alternative to the traditional gas compressing refrigeration due to its high efficiency and environmental friendliness. The unusual negative electrocaloric effect refers to  the adiabatic temperature drops due to application of electric field, in contrast with the normal (positive) ECE, and provides ways to improve the electrocaloric efficiency in refrigeration cycles. However, negative ECE is unusual and requires a clear understanding of microscopic mechanisms. Here, we found unexpected and extensive negative ECE in nonpolar orthorhombic, tetragonal, and cubic phases of halide and oxide perovskite at wide range of temperature by means of first-principle-based large scale Monte Carlo methods. Such unexpected negative ECE originates from the octahedral tilting related entropy change rather than the polarization entropy change under the application of electric field. Furthermore, a giant negative ECE with temperature change of 8.6 K is found at room temperature. This giant and extensive negative ECE in perovskite opens up new horizon in the research of caloric effects and broadens the electrocaloric refrigeration ways with high efficiency.

\end{abstract}

% Halide perovskites have attracted great attention in recent years for the application in photovoltaic cells. Yet their response under the electric field is less studied.
% The adiabatic temperature change under the electric field is konwn as electrocaloric effect, which
% offers promising

%\keywords{}

\maketitle

%%%%%%%%%%%%%%%%%%%%%%%%%%%%%%%%%%%%%%%%%% BEGIN %%%%%%%%%%%%%%%%%%%%%%%%%%%%%%%%%%%%%%%%%%%%

\section{Introduction}

\begin{figure}
  \includegraphics[width=\figurewidthhalf]{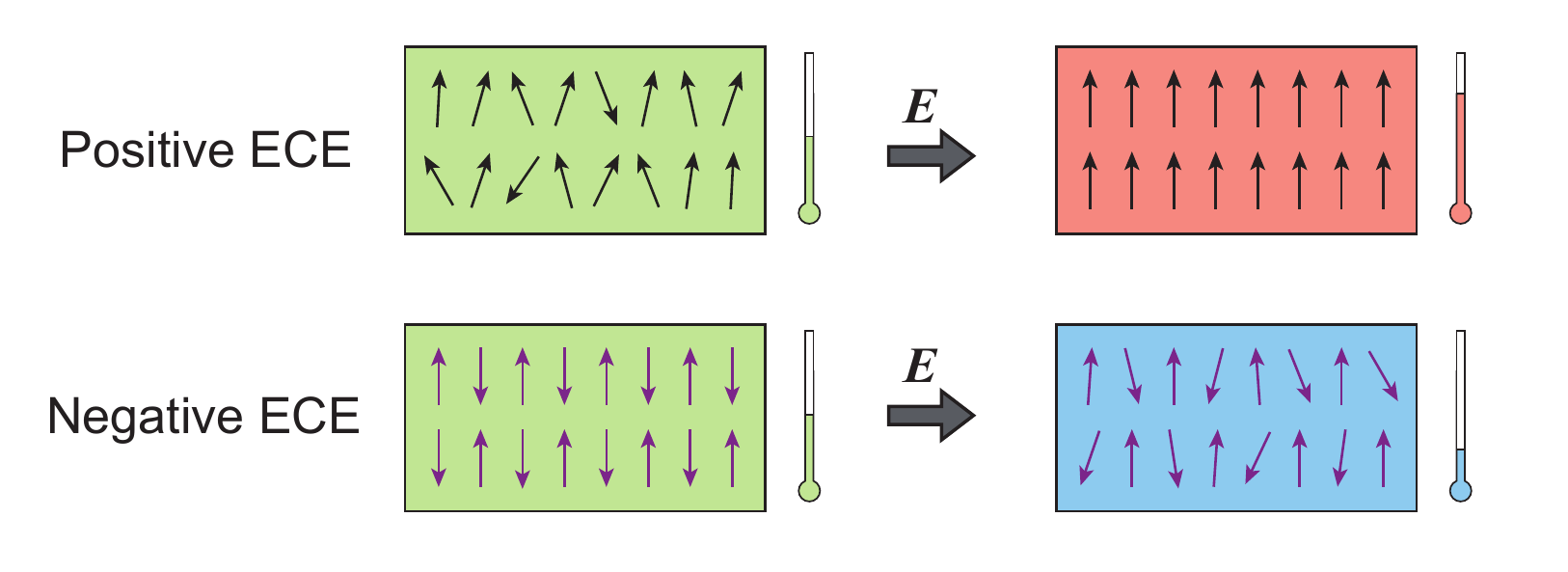}
  \caption{\textbf{Schematics positive and negative ECE.} The top and bottom panels represent the positive and negative ECE, respectively. The arrows in top and bottom panel correspond to local electric dipole and local antiferrodistortive (AFD) octahedral rotation vector, respectively. }
  \label{fig:ECE}
\end{figure}

Refrigeration consumes energy intensively and more than 20\% of the electricity generated in the world is used for cooling \cite{IEAFutureCooling}. Solid state cooling technologies based on the caloric effects display an attractive alternative to the traditional vapor compression cycles due to the higher operating efficiency and zero greenhouse gas emission \cite{Shi2019,moya_caloric_2020,moya_caloric_2014}. Caloric effects are the phenomena of temperature and entropy change induced by electric field (electrocaloric, EC) \cite{Mischenko2006_ECE,ECE_neese_2008,liu_giant_2021}, magnetic field (magnetocaloric) \cite{MCE_franco_2018}, mechanical stresses (elastocaloric \cite{Elasto_tusek_2016}, barocaloric \cite{Baro_li_2019} and twistocaloric \cite{twisto_science.aax6182}). EC based cooling technology is promising in modern microelectronics where millions of electronic units integration to the chip lead to tremendous power dissipation density and high-efficient thermal manage systems is highly required \cite{moore_emerging_2014,ball_computer_2012}. EC effects can be positive (where the entropy decreases and temperature increases with the application of electric field), and negative (where the entropy increases and the temperature decreases with the application of electric field), as shown in Fig. \ref{fig:ECE}. Both of the two EC effects can be used in cooling technologies \cite{Shi2019,NECE_review_grunebohm_origins_2018}. Combining the positive and negative ECE  is proposed to enlarge the EC temperature span and improve the EC efficiency in refrigeration cycles \cite{ponomareva_bridging_2012,NECE_review_grunebohm_origins_2018,NECE_loop_basso_doubling_2014,NECE_loop_ma_enhanced_2016}. Though the positive ECE have been intensively studied in single crystals \cite{moya_giant_2013}, thin films \cite{gao_designed_2021}, ceramic multilayer chips \cite{Nair2019_Nature,Torello2020}, polymer \cite{qian_high-entropy_2021,ECE_neese_2008,ma_highly_2017}, and ceramic/polymer composite \cite{li_thermal_2022,zhang_ferroelectric_2015}, the negative ECE is found only in a few materials, basically complex polar ralxor and antiferroelectric oxdies \cite{NECE_AM_geng_giant_2015,NECE_PZO_vales-castro_origin_2021,ponomareva_bridging_2012,NECE_PMN_PhysRevB.82.134119,NECE_bai_abnormal_2011}. The presenting temperature and magnitude of negative EC coefficient are limited, and its microscopic mechanism is highly required to be revealed.

In addition to oxide perovskites, halide perovskites are also promising candidate materials in cost-effective, high-performance electronics and optoelectronics \cite{CPI_snaith_perovskites_2013,CPI_protesescu_nanocrystals_2015,CPI_swarnkar_quantum_2016}. The temperature change under electric-field will give great impact on performance, degradation and reliability of the materials and devices. However, the response under electric field in halide perovskite and related materials is little studied.

Here, we investigated the EC effect in nonpolar halide (\ce{CsPbI3}) and oxide (\ce{BaCeO3}) perovskite  using first-principles-based large scale Monte Carlo (MC) simulation methods. Perovskite \ce{CsPbI3} and \ce{BaCeO3} can adopt nonpolar structures of orthorhombic, tetragonal, and cubic phases above room temperature. All the nonpolar phases possess negative ECE at small electric-field. The giant negative ECE with temperature change -8.6 K at room temperature is found, which is larger than the typically reported positive ECE with temperature change of +5.5 K \cite{Nair2019_Nature} or negative ECE with temperature change of -5.76 K \cite{NECE_AM_geng_giant_2015}. This large and extensive negative ECE comes from the entropy change related octahedral antiferrodistortive (AFD) motion (see Fig. \ref{fig:phastrans}a) and the strong coupling between AFD and polar motions.
% increase of antipolar entropy and the octahedral tilting entropy under electric-field.

\section{Results}

\subsection{Phenomenological Model}
To understand the ECE in perovskites, the Landau-like model is constructed with four types of non-vanishing order parameters and the homogeneous strain. The four order parameters are (i) AFD motion at M point $\omega_{M,z}=\omega$, (ii) AFD motion at R point $\omega_{R,x}=\omega_{R,y}=\theta$, (iii) antipolar soft mode motion at X point $u_{X,x}=u_{X,y}=s$, and (iv) polar soft mode motion at $\Gamma$ point $u_x=u_y=u$ (which is fully driven by external electric field). According to the symmetry, the components of the homogeneous strain are $\eta_1=\eta_2=\eta_a, \eta_3=\eta_b, \eta_4=\eta_c, \eta_5=\eta_6=\eta_d$. Therefore, using $\omega, \theta, s, u$ and $\eta_a,\eta_b,\eta_c,\eta_d$ as order parameters, the Landau-like model could be written as
\begin{equation}
  \begin{split}
    H_{\text{model}} &=   a_u u^2 + b_u u^4 + a_s s^2 + b_s s^4 \\
    &+ a_\omega \omega^2 + b_\omega \omega^4 + a_\theta \theta^2 + b_\theta \theta^4 \\
    &+ c_1u^2s^2 + c_2\omega^2\theta^2 + ds\omega\theta\\
    &+ f_1\theta^2(u^2+s^2) + f_2\omega^2(u^2+s^2)\\
    &+ E_{\text{elas}}+E_{\text{polar-elas}} + E_{\text{tilt-elas}}\\
    &-2Z^*Eu, \\
  \end{split}
  \label{eq:Het}
\end{equation}
where $\tilde{E}_x=\tilde{E}_y=E$ is the components of external electric field, and $E_{\text{elas}}$ is the energy associated with the homogeneous strain whose details can be found in Ref. \cite{BTO_Zhong1995}, $E_{\text{polar-elas}}$ and $E_{\text{tilt-elas}}$ have following forms
\begin{equation}
  \begin{split}
    E_{\text{polar-elas}} &= B_{1xx}\eta_a(u^2+s^2) + B_{1yy}(\eta_a+\eta_b)(u^2+s^2) \\
    &+ B_{4yz} \eta_c (u^2+s^2),
  \end{split}
\end{equation}
and
\begin{equation}
  \begin{split}
    E_{\text{tilt-elas}} &= \frac{1}{2} C_{1xx}(2\eta_a\theta^2+\eta_b\omega^2) \\
     &+ C_{1yy}(\eta_a \theta^2 + \eta_a \omega^2 + \eta_b \theta^2) \\
     &+ C_{4yz} \eta_c\theta^2.
  \end{split}
\end{equation}
Note that the parameters in this model of Eq. \eqref{eq:Het} can be directly derived from the parameters used in our effective Hamiltonian (see Appendix) which has been confirmed by first principles calculations \cite{CPB_Chen2020}.
For example, $b_u=b_s=4\alpha+\gamma$ and $f_2=2E_{xxyy}$, where $\alpha, \gamma, E_{xxyy}$ are the parameters of effective Hamiltonian calculations \cite{CPB_Chen2020}. Table \ref{tab:param} shows the values of  the parameters in the model of Eq. \eqref{eq:Het}.

\begin{table*}
  \caption{\textbf{Parameters for Landau-like model.} Values are given in atomic units. }
  \label{tab:param}
  \begin{ruledtabular}
    \begin{tabular}{ccccccccc}
        & $a_u$ & $b_u$ & $a_s$ & $b_s$ & $a_\omega$ & $b_\omega$ & $a_\theta$ & $b_\theta$  \\
      \hline
      \ce{CsPbI3}  & -0.000486 &
      0.00270 & -0.000530 & 0.00270 & -0.1017 & 0.9914 & -0.1828 & 4.102 \\
      \ce{PbSc_{0.5}Ta_{0.5}O3} &
      -0.0198 & 0.0519 & -0.00369 & 0.0519 & -0.1981 & 3.777 & -0.4538 & 13.375 \\
      \hline
       & $c_1$ & $c_2$ & $d$ & $f_1$ & $f_2$ & $c_1/a_u$ & $f_1/a_u$ & $f_2/a_u$  \\
      \hline
      \ce{CsPbI3} &  0.0162 & 4.216 & -0.2574 & 0.07393 & 0.07118 & -33.341 & -152.155 &-146.496 \\
      \ce{PbSc_{0.5}Ta_{0.5}O3} &
      0.3112 & 10.482 & 0.1620 & 0.8357 & 0.7556 & -15.714 & -42.196 & -38.153  \\
    \end{tabular}
  \end{ruledtabular}
\end{table*}

To further understand the ECE  from the entropy concept quantitatively, the isothermal entropy change induced by applying electric field is calculated within the model of Eq. \eqref{eq:Het} using Eq. \eqref{eq:dS_general}. The isothermal entropy change can be split into four parts
\begin{equation}
  \begin{split}
    \Delta S=\Delta S_{\text{polar}}+\Delta S_{\text{tilt}}
      +\Delta S_{\text{coup}}+\Delta S_{\text{field}},
  \end{split}
  \label{eq:dS_split}
\end{equation}
where $\Delta S_{\text{polar}}$ is the entropy change only comes from the soft mode, antipolar motion, strain and their coupling
\begin{equation}
  \begin{split}
    \Delta S_{\text{polar}} &= \frac{1}{T} [ a_u\Delta (u^2) + b_u\Delta (u^4) \\
    &+ a_s \Delta (s^2) + b_s \Delta (s^4) + c_1\Delta( u^2s^2) \\
    &+ E_{\text{elas, H}} + E_{\text{polar-elas}}
    ],
  \end{split}
    \label{eq:dS_polar}
  \end{equation}
$\Delta S_{\text{tilt}}$ is the contribution only from AFD motion and its coupling with strain
\begin{equation}
  \begin{split}
    \Delta S_{\text{tilt}} &= \frac{1}{T} [ a_\omega \Delta (\omega^2) + b_\omega \Delta(\omega^4) \\
    &+ a_\theta \Delta(\theta^2) + b_\theta\Delta(\theta^4) + c_2 \Delta(\theta^2\omega^2) \\
    &+ E_{\text{tilt-elas}}
    ],
  \end{split}
  \label{eq:dS_tilt}
\end{equation}
$\Delta S_{\text{coup}}$ is the entropy change comes from the coupling terms between soft mode and AFD
\begin{equation}
  \begin{split}
    \Delta S_{\text{coup}}&=\frac{1}{T}[
      d\Delta (s\theta\omega) + f_1\Delta(\theta^2(u^2+s^2)) \\
      & + f_2\Delta(\omega^2(u^2+s^2))  ],
  \end{split}
  \label{eq:dS_coup}
\end{equation}
$\Delta S_{\text{field}}$ is the contribution directly related to the electric field
\begin{equation}
  \Delta S_{\text{field}}=\frac{2Z^*}{T} \left[
    \int u\mathrm{d} E - \Delta (Eu)
    \right].
\end{equation}
The notation $\Delta (g)$ means the change of value $g$ in the isothermal process. Particularly, the entropy change associated with antipolar motion $\Delta S_{\text{antipolar}}$, which is also a part of $\Delta S_{\text{polar}}$, is given by
\begin{equation}
  \Delta S_{\text{antipolar}}=\frac{1}{T} \left[a_s \Delta (s^2)+b_s \Delta (s^4)\right].
  \label{eq:dS_antipolar}
\end{equation}

\begin{figure*}
  \includegraphics[width=\figurewidthwide]{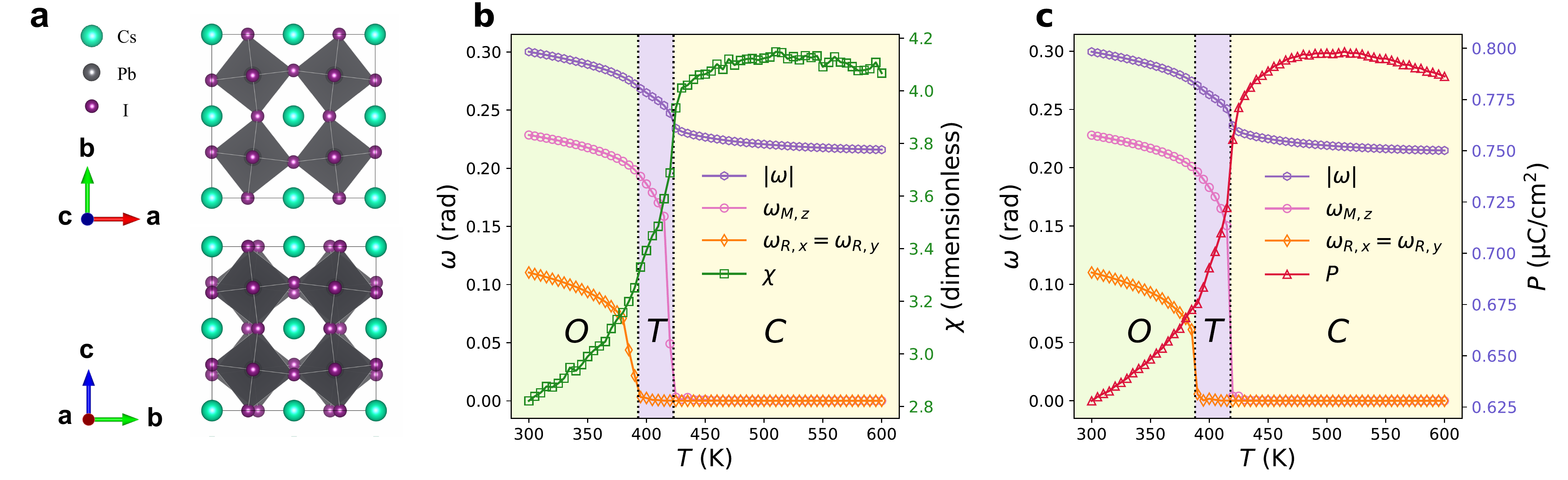}
  \caption{\textbf{ Structures and phase diagram of CsPbI$_3$.}
    (a) Schematics of AFD motions. The top and bottom panels represent the AFD pattern of $\omega_{M,z}$ and $\omega_{R,x}$, respectively.
    (b) AFD vectors $\omega_{M,z}$, $\omega_{R,x}$ ($\omega_{R,y}$) and $|\omega|$, and electric susceptibility $\chi$ as a function of temperature at zero field. (c) AFD vectors $\omega_{M,z}$, $\omega_{R,x}$ ($\omega_{R,y}$) and $|\omega|$, and polarization $P$ as a function of temperature under electric field of 0.43 MV/cm along the pseudocubic [110] direction.  }
  \label{fig:phastrans}
\end{figure*}

\subsection{Structures at finite temperature}

Using first-principles-based effective Hamiltonian and hybrid Monte Carlo metheds (see Appendix A and B), we first investiagted structures and EC properties of halide perovskite structure of \ce{CsPbI3} at finite temperature. Figure \ref{fig:phastrans}b shows  average AFD motions of in-phase tilting at M point ($\omega_{M,z}$) and antiphase tilting at R point ($\omega_{R,x}$) in supercell as a function of temperature (see Fig. \ref{fig:phastrans}a about schematics of such motions, and the corresponding definition in Appendix). At high temperature larger than 420 K, the statistical averages of $\bm\omega_M$ and $\bm\omega_R$ are zero, characterizing a cubic $Pm\bar 3m$ phase (C phase). As the temperature decreases to 420 K, $\omega_{M,z}$ shows a non-zero average value while $\bm\omega_{R}$ is still zero, characterizing a tetragonal $P4/mbm$ phase (T phase) with iodine octahedra tilting pattern $a^0a^0c^+$ (Glazer's notation \cite{Glazer1972}). As for the temperature smaller than 390 K, both $\omega_{R,x}$ and  $\omega_{M,z}$  have non-zero values. The $\omega_{R,x}$ and $\omega_{R,y}$ always have almost identical values below 390 K, characterizing a orthorhombic $Pnma$ phase with tilting pattern $a^-a^-c^+$ (O phase). The phase diagram is  consistent with experimental measurement \cite{CPI_exp_doi:10.1021/acsnano.8b00267} and previous calculations \cite{CPB_Chen2020}.
Note that Figs. \ref{fig:phastrans}b,c also show the average absolute value of AFD $|\omega|$, which possesses a finite value at all investigated temperatures, even for the C phase where both $\bm\omega_M$ and $\bm\omega_R$ vanish, indicating the existence of disordered local AFD in C phase.
%and electric field by performing MC simulations with electric field along pseudocubic [110] direction. Under each magnitude of electric field, the sample is cooled from 600 K with a temperature step of 5 K.

Figure \ref{fig:phastrans}c shows AFD vectors $\omega_{M,z}$ and $\omega_{R,x}$  as a function of temperature under electric field of $\tilde{E}$=0.43 MV/cm.
This phase diagram is very similar to that in the absence of the electric field. The phase transition from C phase to T phase occurs at the temperature of 415 K, and the transition from T phase to O phase occurs at 385 K. Both of the transition temperatures are slightly smaller than that at zero electric field by 5 K.
Interestingly, the polarization shown in Fig. \ref{fig:phastrans}c driven by external electric field \emph{decreases} with the decreasing of temperature. This is  different from prototypical ferroelectrics where the polarization increases with the decrease of temperature and becomes large at lower temperature. This unusual phenomenon of polarization driven by electric field with respect to temperature implies the abnormal temperature response under electric field.
Note that due to the soft mode motion under finite electric field, the symmetry group of the three phases under the application of electric field are no longer \emph{exactly} $Pm\bar 3m$, $P4/mbm$ and $Pnma$, but adopt lower symmetry. However, the notations of C, T and O phase are still used, since the AFD motions at small fields are similar to that in the absence of electric field (see Fig. \ref{fig:phastrans}), and the soft mode motion is rather small.
As the electric field increase above 2.0 MV/cm, no phase transition is observed, and the AFD vectors $\bm{\omega}_R, \bm{\omega}_M$ remain zero in the whole investigated temperature range.

\begin{figure*}
  \includegraphics[width=\figurewidthwide]{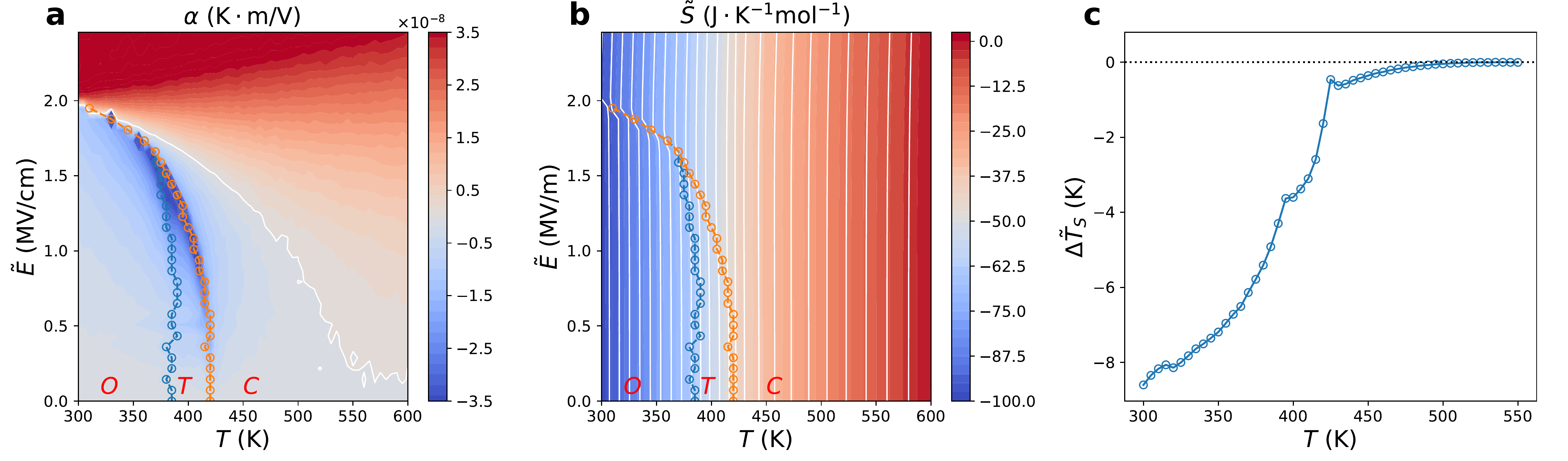}
  \caption{\textbf{Negative ECE in CsPbI$_3$ at wide range of temperature.} (a) Electrocaloric coefficient $\alpha$ as functions of electric field $\tilde{E}$ and temperature $T$. The white solid line denotes the $\alpha=0$ isoline.
    (b) Relative entropy $\tilde{S}$ as functions of electric field $\tilde{E}$ and temperature $T$.
    (c) Maximal negative adiabatic temperature change induced by applying electric field near the phase boundary of brown line shown in panel (a) and (b)  as a function of temperature.
    The blue and brown circles lineas in panels (a) and (b) denote the phase boundaries between O, T, C phases.}
  \label{fig:nece}
\end{figure*}

\subsection{Negative ECE in broad range of temperature}
Figure \ref{fig:nece}a shows the calculated EC coefficient $\alpha$ as functions of temperature $T$ and the applied electric field $\tilde{E}$ along the pseudocubic [110] direction, in which the white solid line is the isoline of zero EC coefficient $\alpha$. The EC coefficient is positive above this isoline (red color), indicating positive ECE, which  generally exists in normal paraelectric materials \cite{ECE_unified_2021}. The EC coefficient is negative in a large region below such isoline (blue color). Interestingly, this negative ECE exists in all the nonpolar phases at a large range of temperature from 300 K to $\simeq 550$ K, very different from that in the previous studies, where negative ECE is typically found in antiferroelectrics phase in a relatively narrow range of temperature \cite{NECE_AM_geng_giant_2015,NECE_PZO_vales-castro_origin_2021}.
Figure \ref{fig:nece}b shows the calculated relative entropy diagram $\tilde{S} (T,\tilde{E})$, where the isentropic lines are depicted by the white solid lines. The adiabatic system evolves in one isentropic line in a reversible  process. On tracing the isentropic lines in the direction of increasing field (i.e. from the bottom to the top in Fig. \ref{fig:nece}b), the left (respectively, right) bending of isentropic lines imply negative (respectively, positive) ECE. Figure \ref{fig:nece}c shows the maximal negative adiabatic temperature change ($\Delta \tilde{T}_S$) as a function of starting temperature, which is the temperature change ($\Delta \tilde{T}_S$) when tracing the isentropic lines in Fig. \ref{fig:nece}b under electric field. The magnitude of negative adiabatic temperature change is large (up to -8.6 K) at low temperature about 300 K, and decreases with the increase of temperature. The negative temperature change is larger than 2 K over a broad range of temperature from 300 K to 415 K.

From Eq. \eqref{eq:Het}, as the external electric field only couples with soft mode motion directly, the enthalpy change induced by change of soft mode under small electric field (where negative ECE is observed) has the form
\begin{equation}
  \begin{split}
    \mathrm{d} H_{\text{model}} &= \{2a_u u + 2(c_1s^2+f_1\theta^2+f_2\omega^2)u + 4b_u u^3 \\
    &+2[B_{1xx}\eta_a + B_{1yy}(\eta_a+\eta_b) + B_{4yz}\eta_c] u - 2Z^*E\} \mathrm{d} u ,
  \end{split}
  \label{eq:Hdu}
\end{equation}
where the last term is directly induced by the electric field and is constant under a given magnitude of electric field, and the third term could be neglected for small $u$. As shown in Table \ref{tab:param}, the parameter $a_u$ for the first term is negative,
indicating the instability of soft mode distortion. For the second term, the large positive $c_1$,  $f_1$ and $f_2$ imply the strong coupling between AFD ($\theta$ and  $\omega$), antipoloar ($s$) and soft mode ($u$) motion, and the antipolar and AFD motion tend to restrain the soft mode distortion. The fourth term typically has small magnitude comparing with the second one.
The values of the same parameters for \ce{PbSc_{0.5}Ta_{0.5}O3} (PST) \cite{PST_PhysRevB.105.054104} which possesses positive ECE are also listed in Table \ref{tab:param}  for comparison. One can see the $f_1/a_u, f_2/a_u, c_1/a_u$ for PST have much smaller magnitudes than those in \ce{CsPbI3}, indicating the coupling between soft mode motions and AFD as well as antipolar motion are \emph{not} strong enough to suppress the soft mode motions in PST, which does not possess negative ECE.

The strong competition between soft mode motion and  AFD motion ($f_1\theta^2 u^2$ and $f_2\omega^2 u^2$) can be confirmed by AFD and polarization in Fig. \ref{fig:phastrans}c. At the small electric filed 0.43 MV/cm, the magnitude of $\omega_{M,z}$ and $\omega_{R,x}$ increase with the decreasing of temperature in T and O phase, while the polarization decreases with the decreasing of temperature from about 500 K.
Therefore, the derivative of polarization $P$ (or the soft mode $u$) with respective to temperature below $\simeq 500$ K at small constant field is positive
\begin{equation}
  \left(\frac{\partial P}{\partial T}\right)_E > 0.
\end{equation}
This contrasts to the fact in prototypical ferroelectrics where the polarization increases with the decrease of temperature. As suggested by Maxwell relations, the EC coefficient could be written as \cite{Shi2019,BZT_PhysRevB.96.014114}
\begin{equation}
  \alpha=-\frac{T}{C_E}\left( \frac{\partial P}{\partial T} \right)_{E},
  \label{eq:ECcoef}
\end{equation}
where $C_E$ is the heat capacity at constant electric field and is positive. It is thus clear that the positive value of $\left(\frac{\partial P}{\partial T}\right)_E$ would result in negative value of $\alpha$ (that is negative ECE). Such analysis implies that the negative ECE below $\simeq 500$ K originates from the strong coupling between AFD and soft mode.

\subsection{Negative ECE in O Phase}
\begin{figure*}
  \includegraphics[width=\figurewidthwide]{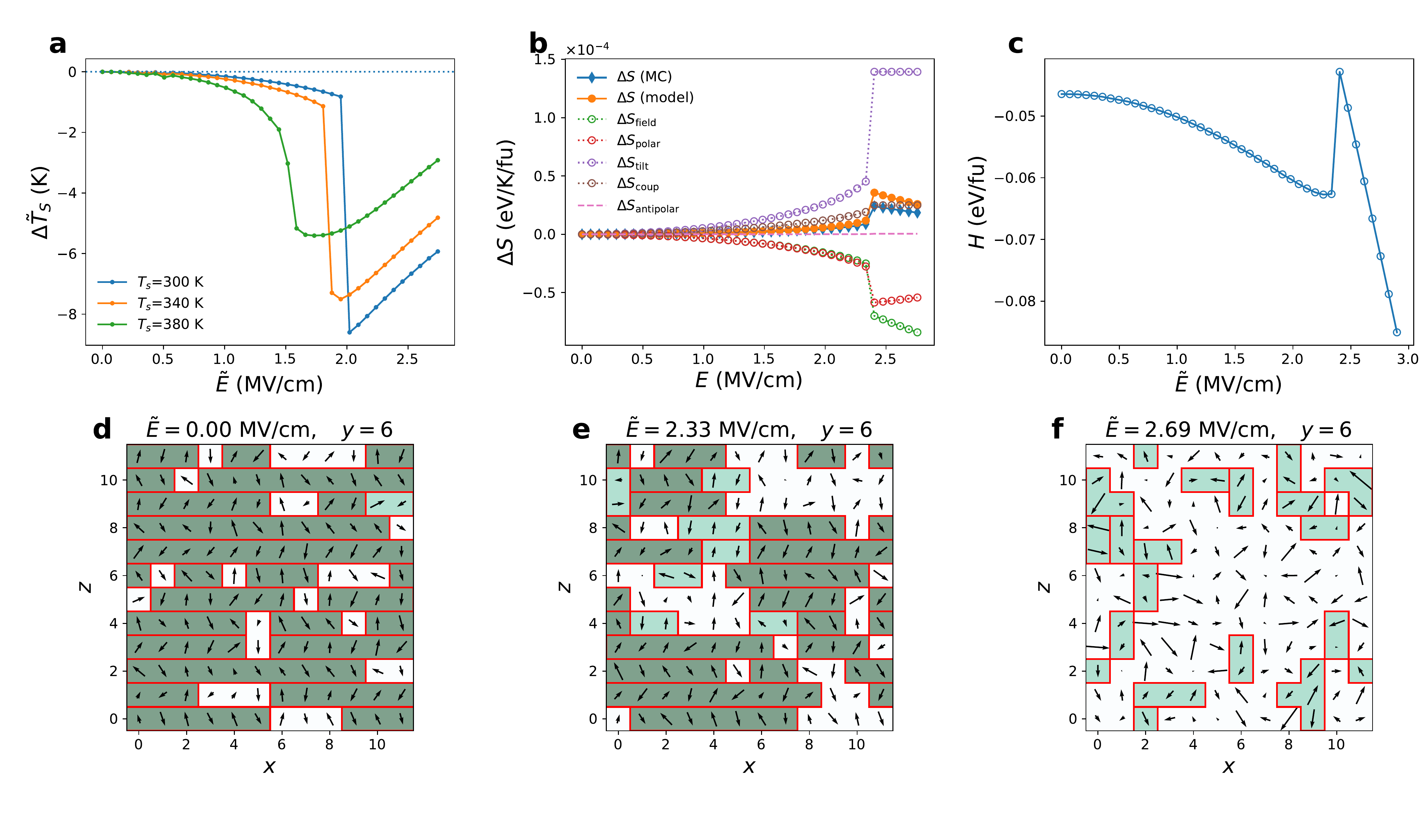}
  \caption{\textbf{Negative ECE in O phase.}
    (a) Adiabatic temperature changes induced by applying electric field starting from the temperature of 300 K, 340 K and 380 K.
    (b) Isothermal entropy changes from the MC simulation [$\Delta S$ (MC), the model of Eq. \eqref{eq:dS_split} [$\Delta S$ (model)] and its contribution from each term, as a function of electric field at 300 K.
    (c) Total enthalpy $H$ as a function of applied electric field along pseudocubic [110] direction at 300 K.
    (d), (e), and (f) Local AFD snapshots at 300 K under electric field of 0.00, 2.33 and 2.69 MV/cm, respectively. Each arrow represents a local AFD vector in an unit cell, and the red lines delimit the \emph{AFD clusters} AFD$_{R,cl}$. Light green color denotes AFD vectors belonging to AFD$_{R,cl}$, dark green backgroud represents percolated AFD$_{R,cl}$.
    \label{fig:Ophase}}
\end{figure*}

%Let us now investigate the ECE in each phase region in detail, starting from the O phase.
The structure of \ce{CsPbI3} is O phase below the temperature 390 K without applying electric field.  The $O$ phase possesses negative ECE (blue color below the white line in Fig. \ref{fig:nece}a) at small electric field. At the electric field above 2.0 MV/cm, the O phase transforms into C phase, and the ECE becomes positive. To further investigate the negative ECE, Fig. \ref{fig:Ophase}a shows the adiabatic temperature change $\Delta \tilde{T}_S$ due to the application of electric field, from the initial temperatures of 300 K, 340 K and 380 K.
For the adiabatic process starting from 300 K, the temperature decreases slightly as the electric field increases from 0 MV/cm, which is consistent with the negative EC coefficient (see Fig. \ref{fig:nece}a). The temperature drop is about -0.82 K at 1.95 MV/cm. As the electric field further increases and passes through the phase boundary between O and C phases, the temperature change drops drastically to -8.6 K. The further increasing of electric field larger than 2.0 MV/cm  results in the increase of the temperature change (positive ECE). The adiabatic process starting from 340 K behaves similarly to that starting from 300 K, possessing a temperature jump at 1.8 MV/cm with an extreme value of -7.5 K. For the starting temperature of 380 K, the temperature change drop occurs at the phase boundary line between O and T phase at lower electric field about 1.5 MV/cm, and exhibits a broader valley at the vicinity of phase boundaries.

To figure out the entropy change in the phenomena of negative ECE in O phase, Fig. \ref{fig:Ophase}b shows the isothermal entropy change computed from the Monte Carlo (MC) simulations by Eq. \eqref{eq:dS_general} [$\Delta S$ (MC)], the model of Eq. \eqref{eq:dS_split} [$\Delta S$ (model)], and the different terms in Eq. \eqref{eq:dS_split}, as a function of electric field  in an isothermal process at 300 K.
The $\Delta S$ (model) curve consistents very well with the $\Delta S$ (MC) curve, indicating the validity of the model of Eq. \eqref{eq:Het} and Eq. \eqref{eq:dS_split}.
The $\Delta S_{\text{antipolar}}$  (Eq. \eqref{eq:dS_antipolar}) is close to zero.
The $\Delta S_{\text{field}}$ and $\Delta S_{\text{polar}}$ curves show negative values in the whole investigated field range, while the $\Delta S_{\text{tilt}}$ and $\Delta S_{\text{coup}}$ display positive values. The positive total entropy change mainly comes from the large $\Delta S_{\text{tilt}}$. Furthermore, the positive $\Delta S_{\text{coup}}$  is the entropy change from coupling between AFD and polar or antipolar motion. Therefore, the positive entropy change $\Delta S$ (model) and the resulting negative ECE come from the AFD related entropy change. The conclusion that the negative ECE is induced by the AFD related entropy change contrasts to the literature \cite{NECE_AM_geng_giant_2015,Shi2019} in which antipolar entropy change is proposed to leading to the negative ECE.
Note that the AFD ($\omega$ and $\theta$) does \emph{not} interplay  with electric field directly (see the full analytical form of the effective Hamiltonian \cite{CPB_Chen2020}), but couples with $u$ by $\omega ^2 u^2$ and $\theta^2u^2$, which leads to the AFD related entropy change due to $u$ (or polarization) responsible for the electric field indirectly.

On the other hand, near 2.4 MV/cm, the $\Delta S_{\text{tilt}}$ and $\Delta S_{\text{coup}}$ terms exhibit a sudden positive jump, leading to a jump of total entropy.
In Fig. \ref{fig:Ophase}c, the enthalpy also exhibits a sudden jump near 2.4 MV/cm. Such jumps of entropy and enthalpy are caused by the phase transition from O phase to C phase driven by the external electric field, which is endothermic in nature.
Further more, such phase transition takes place from an ordered state (with non-vanishing statistical average of $\omega_{M,z}$ and $\omega_{R,x}$ in O phase) to a disordered state (with vanishing order of AFD pattern as well as statistical average of $\omega_{M,z}$ and $\omega_{R,x}$ in C phase). The disordering of AFD motion contributes greatly to the overall entropy, resulting in the large negative ECE near the transition (Fig. \ref{fig:Ophase}a).
This  endothermic phase transition induced negative ECE also occurs in \ce{PbZrO3} \cite{NECE_PZO_vales-castro_origin_2021}.

In order to gain a microscopic understanding of the AFD entropy change, the \emph{local clusters} AFD$_{R,cl}$ (or AFD$_{M,cl}$) are defined to describe the local ordering of the AFD pattern. At finite temperature under electric field, the AFDs are not fully ordered, but partially ordered.
The AFD$_{R,cl}$ (respectively, AFD$_{M,cl}$) is defined as the local clusters in which the AFD vector in each unit cell $\bm \omega_R (i)=(-1)^{n_x(i)+n_y(i)+n_z(i)} \bm \omega_i$ [respectively, $\bm \omega_M (i)=(-1)^{n_x(i)+n_y(i)}$] are nearly parallel to each other.
The \emph{local clusters} AFD$_{K,cl}$ (where $K$ corresponds to $R$ or $M$ here) are practically identified by comparing directions of $\bm\omega_K(i)$ with their nearest neighbors \cite{BZT_filmPhysRevLett.111.247602}, two neighbor AFD vectors are considered to belong to the same cluster if the cosine of the angle between them is larger than 0.85. The \emph{local clusters} AFD$_{K,cl}$ can be described by
size $\langle s\rangle$ (which is the number of units in the cluster) and
the average magnitude of AFD vector in the cluster $\omega_{K, \alpha}^{\text{clus}}$ defined by
\begin{equation}
  \omega_{K, \alpha}^{\text{clus}}=\frac{1}{N_t} \sum_{j} |\omega_{R,\alpha}(j)|,
\end{equation}
where $\alpha$ is the cartesian direction index, $j$ sums over all the unit cells that belongs to a local AFD cluster with size larger than 1, and $N_t$ is the number of such unit cells in the simulated supercell.
Note that the cluster can propagates from one side of the supercell to its opposite side, which is the so-called \emph{AFD percolation}, the cluster of \emph{AFD percolation} has infinite size under periodic boundary conditions.

As shown in Figs. \ref{fig:Ophase}d, e and f, the size $\langle s\rangle$ of the \emph{AFD cluster} decreases when the electric field increase. Almost all unit cells belong to the clusters of percolation in the absence of field, some AFD vector do not belong to the AFD cluster and percolation at electric field of 2.33 MV/cm, and there is no percolation and most AFD vector do not belong to clusters at the electric field of 2.69 MV/cm. The ordered AFD vectors in the absence of field become disordered at high electric field, and the AFD entropy increases with the increase of electric field.

\begin{figure*}
  \includegraphics[width=\figurewidthwide]{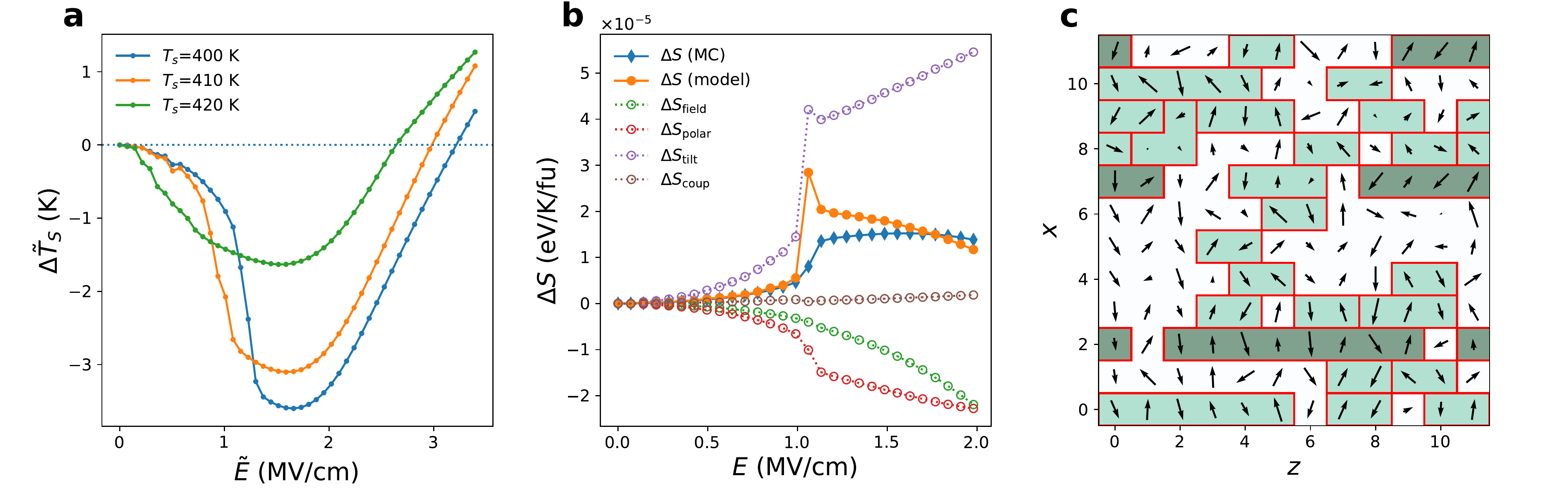}
  \caption{\textbf{Negative ECE in T phase.}
    (a) Adiabatic temperature changes induced by applying electric field starting from 400, 410 and 420 K in T phase.
    (b) Isothermal entropy change induced by applying electric field computed directly from MC results [$\Delta S$ (MC)], the simplified model [$\Delta S$ (model), Eq. \eqref{eq:dS_split}], and the contribution from each term (defined in text), as a function of electric field at 400 K.
    (c) Schematics of \emph{AFD cluster} AFD$_{R,cl}$ taken from MC simulation at 410 K in the absence of field. The arrows denote the local AFD vectors, and the red lines delimit the \emph{AFD clusters} AFD$_{R,cl}$. Light green color denotes AFD vectors belonging to AFD$_{R,cl}$, dark green background represents percolated AFD$_{R,cl}$.
    }
  \label{fig:Tphase}
\end{figure*}

\subsection{Negative ECE in T phase}

The ECE in T phase is then investigated.
% We then investigate the ECE in T phase.
As shown in Fig. \ref{fig:nece}a, the T phase possesses negative ECE at low electric field. The EC coefficient exhibits large negative value at the phases boundary of O, T, and C phases near the temperature of 390 K under electric field of 1.4 MV/cm.  This large negative EC coefficient (dark blue color in Fig. \ref{fig:nece}a) exists in a wide range of electric field from 1.3 MV/cm to 2.0 MV/cm.
Figure \ref{fig:Tphase}a shows the adiabatic temperature change as a function of electric field starting from the temperatures of 400 K, 410 K and 420 K, where the structure of \ce{CsPbI3} is T phase at zero electric field.
At low field, the larger the starting temperature is, the faster adiabatic temperature change ($\Delta\tilde{T}_S$)  drops, and the smaller the maximal negative adiabatic temperature change is. For starting temperature of 400, 410 and 420 K, the maximal negative adiabatic temperature changes are -3.6 K at about 1.7 MV/cm, -3.1 K at about 1.6 MV/cm, and -1.6 K at about 1.5 MV/cm, respectively.
Note that the  maximal adiabatic temperature change terrace $\Delta \tilde{T}$ presents at larger range of temperature than that of O phase (see Fig. \ref{fig:Ophase}a), which is consistent with Fig. \ref{fig:nece}a.

To understand the unexpected negative ECE in T phase, the entropy change is analyzed with the model proposed in Eq. \eqref{eq:Het} and Eq. \eqref{eq:dS_split}. In T phase the antipolar motion $s$ is zero and the antipolar entropy change $\Delta S_{\text{antipolar}}$ is zero. Note that the AFD at R point ($\theta$) is in principle zero in the whole supercell of T phase. However, considering the the AFD may be partially ordered and form local AFD cluster \cite{CPI_dyn_https://doi.org/10.1002/adfm.202106264}, here the $\theta$ is numerically defined as the average value of $\omega_R$ in AFD$_{R,cl}$ clusters, i.e. $\omega_{R,x}^{\text{clus}}$. Figure \ref{fig:Tphase}c shows the snapshot taken from MC simulation at 410 K at the absent of electric field, confirming the existence of AFD$_{R,cl}$ clusters in macroscopic T phase.
It is numerically found such treatment of $\theta$ is essential for our model to reproduce the entropy change from MC calculations correctly.
Figure \ref{fig:Tphase}b shows the isothermal entropy change computed from MC simulation [$\Delta S$ (MC)], the model proposed in Eq. \eqref{eq:dS_split} [$\Delta S$ (model)] and its contribution from each term, as a function of electric field at 400 K. The $\Delta S$ (model) is consistent very well with the $\Delta S$ (MC) at low field and also close to each other at high field.
Similar to the case of O phase, the $\Delta S_{\text{field}}$ and $\Delta S_{\text{polar}}$ terms possess large negative values. The positive contribution mainly comes from $\Delta S_{\text{tilt}}$. It is thus clear that the fundamental origin of NECE in T phase is also the tilting entropy.
Note that the contribution from $\Delta S_{\text{coup}}$ is rather small in T phase, in contrast with that in O phase, where the $\Delta S_{\text{coup}}$ has significantly positive value (see Fig. \ref{fig:Ophase}b). It is numerically found that such difference stems from the trilinear coupling term $d\Delta (s\theta\omega)$ of Eq. \eqref{eq:dS_coup} in O phase. In fact, the trilinear term is the main contribution to $\Delta S_{\text{coup}}$ in O phase. However, such term vanishes in T phase, since the antipolar motion $s$ is zero in T phase, leading to the small total value of $\Delta S_{\text{coup}}$.

\subsection{Electrocaloric Response in C Phase}

\begin{figure*}
  \includegraphics[width=\figurewidthwide]{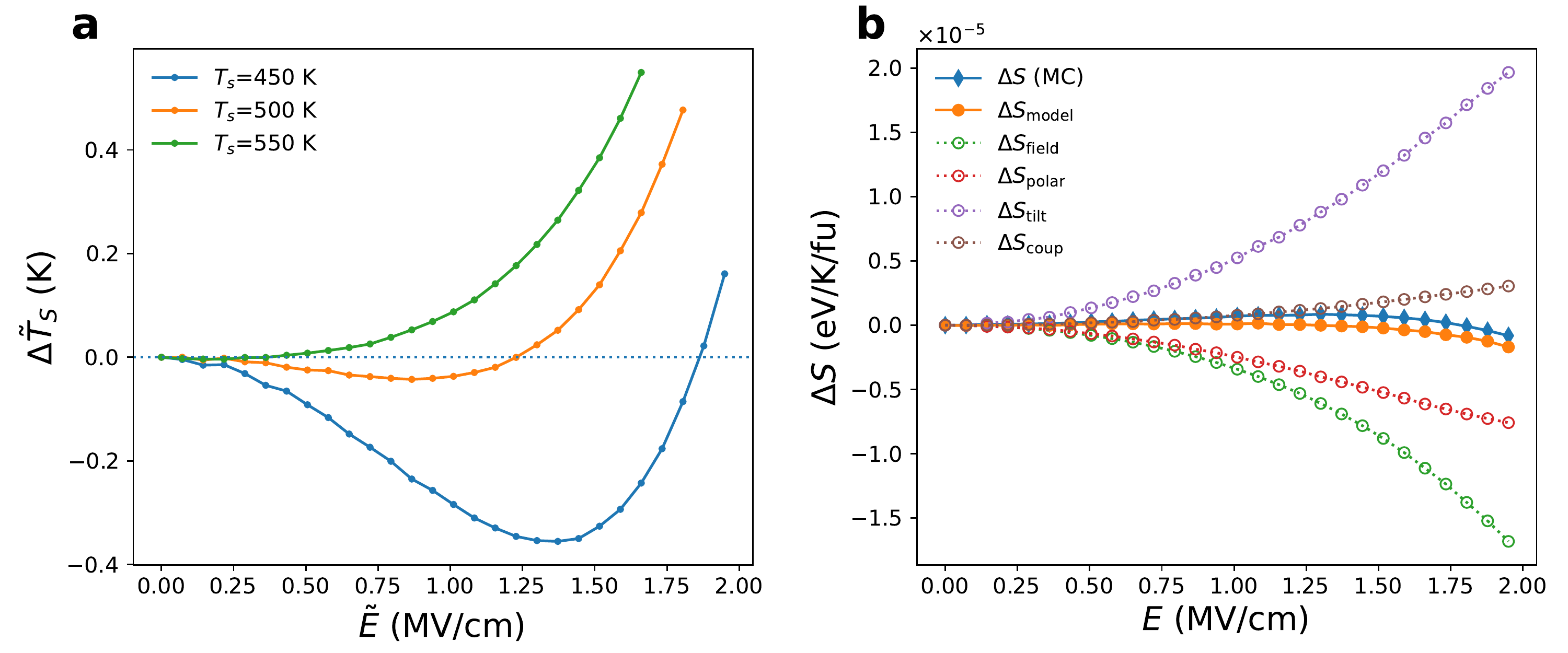}
  \caption{\textbf{Electrocaloric response in C phase.}
    (a) Adiabatic temperature change induced by applying electric field starting from the temperature of 450 K, 500 K and 550 K in C phase.
    (b) Isothermal entropy change computed from MC simulation [$\Delta S$ (MC)], simplified model [$\Delta S$ (model), Eq. \eqref{eq:dS_split}] and its contribution terms, as a function of electric field at 450 K.
    }
  \label{fig:cubic}
\end{figure*}

C phase of \ce{CsPbI3} also presents negative ECE. As shown in Fig. \ref{fig:cubic}a, the adiabatic temperature change starting from 450 K and 500 K can be negative at low electric field.
The negative adiabatic temperature starting from 450 K reaches the maximum of -0.36 K at the electric field of 1.37 MV/cm. This maximum of negative ECE occurs at the critical electric field where $\alpha=0$ shown in in Fig. \ref{fig:nece}a. The negative ECE become weaker when the starting temperature increases, characterized by the decreasing absolute values of temperature change under electric field. The negative ECE vanishes completely at the  starting temperature of 550 K, where the adiabatic temperature change is positive for all investigated magnitudes of electric field, consistent with the fact that the $\alpha$ is almost always positive at temperature larger than 550 K. The ECE becoming positive at the temperature larger than 550 K is also consistent with the fact that $\left(\frac{\partial P}{\partial T}\right)_E$ is negative at low electric field (see Fig. \ref{fig:phastrans}c) which leads to the positive $\alpha$ from Eq. \eqref{eq:ECcoef}.

Similar to the local cluster AFD$_{R,cl}$ in T phase, there are partially ordered AFD of local clusters AFD$_{R,cl}$ and AFD$_{M,cl}$ in C phase.  The average values of $\omega_{R}$ (respectively, $\omega_{M}$) inside the AFD$_{R,cl}$ (respectively, AFD$_{M,cl}$) clusters are used as $\theta$ (respectively, $\omega$) in the models of Eq. \eqref{eq:Het} and Eq. \eqref{eq:dS_split} to analyze the entropy change.
Figure \ref{fig:cubic}b shows the isothermal entropy change computed from MC simulation [$\Delta S$ (MC)], the model in Eq. \eqref{eq:dS_split} [$\Delta S$ (model)] and its contribution from different terms, as a function of electric field at 450 K. Similar to T phase, the $\Delta S_{\text{field}}$ and $\Delta S_{\text{polar}}$ are negative,  $\Delta S_{\text{coup}}$ is almost zero, $\Delta S_{\text{tilt}}$ is positive and very large. Clearly, it is $\Delta S_{\text{tilt}}$ that mainly results in the negative ECE in C phase, similar to that in T phase where the large positive tilting entropy change comes from the strong coupling between AFD and soft mode motions in local clusters under electric field.

\section{Discussion}

All the nonpolar phases of O phase ($a^-a^-c^+$), T phase ($a^0a^0c^+$) and C ($a^0a^0a^0$) phase in \ce{CsPbI3} possess negative ECE, originating from the AFD (highly ordered or partly ordered) entropy change. The result of AFD entropy changes resulting in negative ECE is different from the hypothesis in which antipolar entropy change gives rise to the negative ECE \cite{NECE_AM_geng_giant_2015,Shi2019}. In \ce{CsPbI3}, the antipolar entropy change is almost zero in O phase, and the there is no macroscopic antipolar motion in T and C phases. Our results of negative ECE in the paraelectric phase of T and C phases are in contrast with the calculations of the unified perturbation model \cite{ECE_unified_2021} in which the paraelectric compounds possess \emph{positive} ECE. This is because of the parameter of the AFD cluster and its coupling with soft mode, which leads to the the dielectric susceptibility $\chi$ decreases with the decreasing of temperature (Fig. \ref{fig:phastrans}a), in contrast with the normal Curie-Weiss law \cite{ECE_unified_2021}.

\begin{figure}
  \includegraphics[width=\figurewidthhalf]{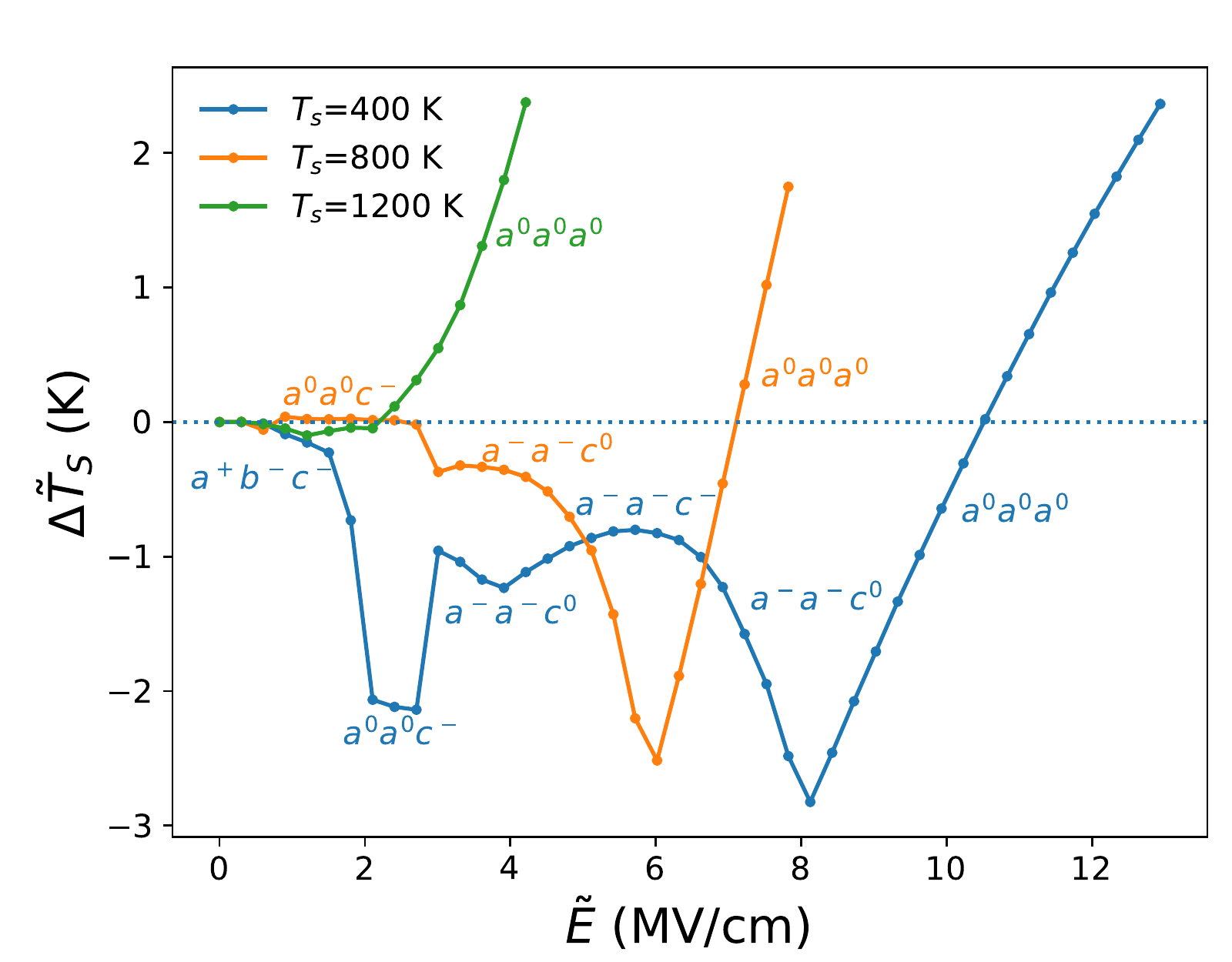}
  \caption{\textbf{Negative ECE in \ce{BaCeO3}}. Adiabatic temperature change of  \ce{BaCeO3} as a function of electric field starting from 400 K (in $Pbnm$ O phase), 800 K (in $I4/mcm$ T phase) and 1200 K (in $Pm\bar 3m$  C phase). The AFD patterns of structures on electric field are labeled on the corresponding curves.}
  \label{fig:BCO}
\end{figure}

Such negative ECE comes from the AFD entropy change in nonpolar phases can also exist in other perovskites, including both halide and oxide peroskites. We take \ce{BaCeO3} as an example to investigate the ECE in oxide perovskites.  As suggested in previous measurements \cite{BCO_PhysRevB.82.014113,BCO_JAmCerSoc2008} and first-principles-based calculations \cite{BCO_PhysRevB.104.174102}, \ce{BaCeO3} possesses paraelectric $Pm\bar 3m$ phase, nonpolar tetrahedral $I4/mcm$ phase (with AFD pattern $a^-a^0a^0$) and antipolar orthorhombic $Pbnm$ phase (with AFD pattern $a^-a^-c^+$) as the temperature decreases from 1200 K to room temperature in a monodomain form.
The adiabatic temperature change are calculated for different phases of \ce{BaCeO3} \cite{BCO_PhysRevB.104.174102} as shown in Fig. \ref{fig:BCO}. The structure phase is O phase ($a^-a^-c^+$) at the starting temperature of 400 K, and T phase ($a^-a^0a^0$) at the strating temperature of 800 K, C phase ($a^0a^0a^0$) at the starting temperature of 1200 K. All the phases exhibit negative ECE at certain small electric field range.
% Detail analysis shows that these negative ECE maily come from the AFD entropy change, very similar to that in halide perovskite \ce{CsPbI3}.  % We do not have evidence for this declaration ...

The novel phenomena of negative ECE  is found in a very broad range of temperature in nonpolar phases of O, T and C phases for halide and oxide perovskites. These negative ECE is induced by the octahedra AFD related entropy change under the application of electric field and by the strong coupling between AFD and polarization in the nonpolar phases.
We thus expect that the presently determined negative ECE would be measured as the important thermal control in the perovskite application in electronic, photoelectric and photovoltaic devices.

%The NECE originates from strong coupling between AFD and soft mode motions is proposed here for the first time, to the best of our knowledge. Such type of NECE could be relatively strong in a broad range of electric field (see the comparason between Fig. \ref{fig:Ophase}a and \ref{fig:Tphase}a) with smooth in nature, in contrast with the NECE associated with first-order phase transition driven by electric field (see Fig. \ref{fig:Ophase}a, and \cite{NECE_PZO_vales-castro_origin_2021}).

% To the best of our knowledge, this is the first time that NECE originates from such coupling mechanism is reported, and our work opens a new class of NECE materials, with the potential to possess relatively large NECE in a broad range. We thus hope our work not only leads to a better understanding of NECE in dielectrics, but also open up a new area for searching the NECE materials.

% \end{document}

\appendix
\section{Effective Hamiltonian}
We use the effective Hamiltonian methods developed in Ref. \cite{CPB_Chen2020}, along with the parameters computed from first-principle calculations.
The degrees of freedom of effective Hamltonian are local soft mode vectors $\lbrace \bm {u}_i\rbrace$ of each 5-atom unit cell, local displacement vectors $\lbrace \bm{v}_i\rbrace$ related to inhomogeneous strain, pseudo vectors $\lbrace \bm{\omega}_i\rbrace$ representing the rotation of iodine octahedra [also known as antiferrodistortive (AFD) motions], and homogeneous strain tensor $\eta_H$. The total energy of the effective Hamiltonian has two main terms
\begin{equation}
  E_{\text{tot}} = E_{\text{dipole}} (\lbrace \bm u_i\rbrace, \lbrace\bm v_i\rbrace, \eta_H) +
  E_{\text{AFD}}(\lbrace \bm u_i\rbrace,\lbrace \bm v_i\rbrace,\lbrace\bm\omega_i\rbrace, \eta_H),
  \label{eq:Etot}
\end{equation}
where $E_{\text{dipole}}$ is the energy of local soft modes, strains and their coupling, including five main terms \cite{BTO_Zhong1995}, namely, the quartic soft mode self energy, the quadratic soft mode long range dipolar energy, the quadratic soft mode short range interaction energy (up to the third nearest neighbor), the elastic energy (including both homogeneous and inhomogeneous contributions), and the interaction between soft mode and strain (quadratic in soft mode and linear in strain); the $E_{\text{AFD}}$ is the energy of AFD motions and their coupling with strains and soft modes, including the quartic AFD onsite energy, the quadratic and quartic AFD short range interaction energy (up to the first neighbor), the interaction between AFD and strain (quadratic in AFD and linear in strain), and the biquadratic and trilinear interaction between soft mode and AFD motion. The interaction coefficient are greatly simplified due to the symmetry. The complete analytical form of the effective Hamiltonian is proposed in Ref. \cite{CPB_Chen2020}.
%Note that the model Hamiltonian proposed in Eq. \eqref{eq:Het} in the main text is the simplified version of the effective Hamiltonian described above. More precisely, the variables used in Eq. \eqref{eq:Het} are the macroscopic order parameters, instead of the local modes in each unit cell used here.
An additional term $-Z^* \sum_i \bm u_i \cdot \bm E$ is added in $E_{\text{tot}}$ to consider the effect of external electric field \cite{BZT_2013PhysRevLett.110.207601}, where the summation of $i$ runs over all the unit cells in the simulated supercell, $Z^*$ is the Born effective charge of the soft mode, and $\bm E$ is the external electric field. Note that $\tilde{\bm E}$ is the rescaled field according the MC calculations and the measurements \cite{BZT_PhysRevB.96.014114,AFE_Xu2017,PST_PhysRevB.105.054104}.

%Note that the considering electric field in atomic schemes are typically overestimated by 1 to 2 orders of magnitudes due to the fact of Landauer paradox \cite{BZT_PhysRevB.96.014114,AFE_Xu2017,PST_PhysRevB.105.054104}, the electric field is rescaled linearly by $\tilde{\bm E}=c\bm E$, where $\bm E$ and $\tilde{\bm E}$ are the original and rescaled field, and $c$ is 0.2 here which gives perfectly consistence with measurement for the calculation of ECE in PST \cite{PST_PhysRevB.105.054104}. %The quantities related to the rescaled electric field are marked with a tilde throughout this manuscript.

The perovskite structures are typically simulated by $12\times 12\times 12$ supercells (corresponding to 8640 atoms) or large $32\times 32\times 32$ supercells (corresponding to 163840 atoms) with Monte Carlo (MC) simulations. The following quantities are computed: (i) The AFD at R point characterizing anti-phase tilting, defined as $\bm{\omega}_R = \frac{1}{N} \sum_i  \bm\omega_i (-1)^{n_x(i)+n_y(i)+n_z(i)}$, where the summation of $i$ runs over all of the $N$ five-atom perovskite unit cells in the simulated supercell, $\bm\omega_i$ is the AFD vector of unit cell located at
$a_0(n_x(i)\hat{\bm x}+n_y(i)\hat{\bm y}+n_z(i)\hat{\bm z})$,
$a_0$ is the lattice constant of the five-atom perovskite unit cell, and $\hat{\bm x}, \hat{\bm y}, \hat{\bm z}$ are the unit vectors along the pseudocubic $[100]$, $[010]$ and $[001]$ directions, respectively; (ii) the AFD at M point characterizing the in-phase tilting, defined as $\bm{\omega}_{M} = \frac{1}{N} \sum_i \bm\omega_i (-1)^{n_x(i)+n_y(i)}$;
(iii) the average absolute value of local AFD vectors $|\omega|=\frac{1}{N} \sum_i |\bm \omega_i|$;
(iv) the average soft mode motion $\bm u= \frac{1}{N} \sum_i \bm u_i$; (v) the polarization $\bm P=\frac{1}{\Omega} Z^* \bm u$, where $\Omega$ is the volume of the simulated supercell; and (vi) the antipolar soft mode motion at X point $\bm{u}_X=\frac{1}{N} \sum_i \bm u_i (-1)^{n_z(i)}$.
%Figure \ref{fig:phastrans}a shows the schematics of the AFD motions. Note that for the simplicity, we performed coordinate transformation to ensure the maximum the average $\bm{\omega}_M$ is along the $z$ axis at low temperature.

The dielectric susceptibility tensor $\chi$ shown in Fig. \ref{fig:phastrans}b and the EC coefficient $\alpha$ shown in Fig. \ref{fig:nece}a are computed from the MC simulations with the cumulant formula \cite{BZT_2012PhysRevLett.108.257601,BZT_PhysRevB.96.014114,BFO_ECPhysRevB.103.L100102}.

The isothermal entropy change is calculated using the numerical method developed in Ref. \cite{PST_PhysRevB.105.054104} as
\begin{equation}
  \Delta S=\frac{1}{T} \left(
  \Delta H + \int \Omega \bm P \cdot \bm E
  \right),
  \label{eq:dS_general}
\end{equation}
where $H$ is the enthalpy calculated by
\begin{equation}
    H=E_{\text{tot}} + \frac{21}{2} Nk_BT +p\Omega -\Omega \bm E\cdot \bm P,
  \label{eq:enthalpy}
\end{equation}
where $N$ is the number of unit cells in the supercell, $k_B$ is the Boltzmann constant, $T$ is the temperature, $E_{\text{tot}}$ is the energy of effective Hamiltonian provided in Eq. \eqref{eq:Etot}, $p$ is the external pressure ($p=0$ in this work), $\bm E$ is the external electric field, and $\bm P$ is the polarization of the supercell.
The entropy change as a function of temperature with absent of electric field is computed by
\begin{equation}
  \begin{split}
      \Delta S|_{\bm{E}=0} &= \int \frac{1}{T} \mathrm dH \\
      &=\Delta \left( \frac{H}{T} \right) + \int \frac{H}{T^2}\mathrm{d}T.
  \end{split}
  \label{eq:dS_ZF}
\end{equation}
The relative entropy diagram $\tilde{S}(T,\tilde{E})$ is thus determined from Eqs. \eqref{eq:dS_general} and \eqref{eq:dS_ZF}, as shown in Fig. \ref{fig:nece}b.
% The adiabatic temperature change $\Delta \tilde{T}_S$ is calculated by tracing the isentropic line in the relative entropy $\tilde{S} (T,\tilde{E})$ diagram (see Ref. \cite{PST_PhysRevB.105.054104}).
The adiabatic temperature change $\Delta \tilde{T}_S$ is calculated from the relative entropy diagram. More precisely, the adiabatic temperature change $\Delta T_S (T_{\text{start}},\tilde{E})$ from certain temperature $T_{\text{start}}$ induced by applying certain electric field $\tilde{E}$ is determined by finding the temperature $T_1$ at the given field $\tilde{E}$ that satisfies $\tilde{S}(T_1,\tilde{E}) = \tilde{S} (T_{\text{start}},0)$, and the temperature change is then calculated by $\Delta \tilde{T}_S(T_{\text{start}},\tilde{E})=T_1-T_{\text{start}}$.

\section{Hybrid Monte Carlo Simulation}
Hybrid Monte Carlo  (HMC) algorithm \cite{HMC_duane_hybrid_1987,HMC_mehlig_hybrid_1992,HMC_prokhorenko_large_2018} is implemented for large scale simulations.
In each Monte Carlo sweep (MCS) of the HMC simulations, a new trial configuration of $\lbrace \bm u_i\rbrace, \lbrace \bm v_i\rbrace$ and $ \lbrace \bm \omega_i\rbrace$ is generated by performing microcanonical molecular dynamics (MD) simulation for a short period with random initial momenta, while the homogeneous strain $\eta_H$ is practically updated using the standard Metropolis algorithm as in Ref. \cite{BTO_Zhong1995}.

Technically, the long range dipole energy
\begin{equation}
  E_{\text{dpl}} = \sum_{ij\alpha\beta} Q_{\alpha\beta} (\bm{r}_i-\bm{r}_j) u_{\alpha} (\bm{r}_i) u_{\beta} (\bm{r}_j)
  \label{eq:dpl}
\end{equation}
and forces associated with it are computed in the reciprocal space with the help of fast Fourier transformation algorithm, as in Refs. \cite{HMC_prokhorenko_large_2018,MD_waghmare_ferroelectric_2003,MD_nishimatsu_fast_2008}. Such treatment reduces the overall computation complexity of one MCS from $O(N^2)$ in standard Metropolis algorithm to $O(N\log N)$. Moreover, in the MD simulations, the degrees of freedom of all unit cells are updated simultaneously, in contrast with Metropolis MC simulation, in which the degrees of freedom are updated in sequence. Such simultaneous update makes it easy to run the HMC simulations \emph{in parallel}, especially on shared memory architechtures \cite{HMC_prokhorenko_large_2018}. The reduced complexity and ability to run in parallel makes it possible to perform simulations on large supercells efficiently.

Moreover, computational efficiency tests were performed for the $E_{\text{elas,I}}$ in Ref. \cite{BTO_Zhong1995} and the following terms (in notations of Refs. \cite{BTO_Zhong1995,CPB_Chen2020})
\begin{align}
  E_{\text{mode-strain}} & = \frac{1}{2} \sum_i \sum_{l\alpha\beta}  B_{l\alpha\beta}
  \eta_{I,l} (\bm r_i) u_\alpha (\bm r_i) u_\beta (\bm r_i),  \label{eq:mode_strain}              \\
  E_{\text{AFD-strain}}  & = \frac{1}{2} \sum_i \sum_{l\alpha\beta}  C_{l\alpha\beta}
  \eta_{I,l} (\bm r_i) \omega_\alpha (\bm r_i) \omega_\beta (\bm r_i),                            \\
  E_{\text{mode-AFD,1}}  & = \sum_{ij} \sum_{\alpha\beta} D_{ij\alpha\beta}
  u_\alpha (\bm r_j) \omega_\alpha (\bm r_i) \omega_\beta (\bm r_i),                              \\
  E_{\text{mode-AFD,2}}  & = \sum_{ij} \sum_{\alpha\beta\gamma\delta} E_{\alpha\beta\gamma\delta}
  \omega_{\alpha}(\bm r_i)\omega_{\beta} (\bm r_i) u_\gamma (\bm r_j) u_\delta (\bm r_j)
\end{align}
where the summation of $i$ runs over all the unit cells in the supercell, the summation of $j$ runs over several neighbor cells around $i$, $\lbrace\eta_I\rbrace$ is the inhomogeneous strain which is computed from $\lbrace \bm v_i\rbrace$ \cite{BTO_Zhong1995}, $\alpha,\beta,\gamma,\delta$ are cartesian directions, and $l=1,2,\cdots,6$ is the Voigt notation index.
Such tests show that it is more efficient to compute these terms (as well as their associated forces) in the reciprocal space, although they have in principle complexity of $O(N)$ calculating in the real space.

\begin{acknowledgments}
  X.M., Y.Y. and D.W. thank the National Key R$\&$D Programs of China (grant NOs. 2020YFA0711504, 2022YFB3807601), the National Science Foundation of China (grant NOs. 12274201, 51725203, 51721001, 52003117 and U1932115) and the Natural Science Foundation of Jiangsu Province(grant NO. BK20200262). We are grateful to the HPCC resources of Nanjing University for the calculations.
\end{acknowledgments}

%%%%%%%%%%%%%%%%%%%%%%%%%%%%%%%%%%%%%%%%%%% END %%%%%%%%%%%%%%%%%%%%%%%%%%%%%%%%%%%%%%%%%%%%%

\bibliography{nece}

\end{document}